\documentclass[aps,prl,twocolumn,noshowpacs,superscriptaddress,groupedaddress]{revtex4}  
\usepackage{graphicx}  
\usepackage{dcolumn}   
\usepackage{bm}        
\usepackage{amssymb}   
\usepackage{amsmath}
\usepackage{color}
\usepackage[colorlinks=true, pdfstartview=FitV, linkcolor=blue, citecolor=blue, urlcolor=blue]{hyperref} 
\usepackage{epstopdf}

\usepackage{appendix} 

\newcommand{\affilPhotonics}{Photonics Laboratory, ETH Z\"urich, CH-8093 Z\"urich, Switzerland.}
\newcommand{\affilITP}{Institute for Theoretical Physics, ETH Z\"{u}rich, CH-8093 Z\"urich, Switzerland.}
\newcommand{\affilIFP}{Laboratory for Solid State Physics, ETH Z\"{u}rich, CH-8093 Z\"urich, Switzerland.}

\begin{document}

\preprint{}

\title{Rapid flipping of parametric phase states}

\author{Martin Frimmer}\affiliation{\affilPhotonics}\homepage{http://www.photonics.ethz.ch}
\author{Toni L. Heugel}\affiliation{\affilITP}
\author{\v{Z}iga Nosan}\affiliation{\affilIFP}
\author{Felix Tebbenjohanns}\affiliation{\affilPhotonics}
\author{David H\"alg}\affiliation{\affilIFP}
\author{Abdulkadir Akin}\affiliation{\affilIFP}
\author{Christian L. Degen}\affiliation{\affilIFP}
\author{Lukas Novotny}\affiliation{\affilPhotonics}
\author{R. Chitra}\affiliation{\affilITP}
\author{Oded Zilberberg}\affiliation{\affilITP}
\author{Alexander Eichler}\email{eichlera@phys.ethz.ch}\affiliation{\affilIFP}\email{eichlera@phys.ethz.ch}
%
%
\date{\today}


\begin{abstract}
 We experimentally demonstrate flipping the phase state of a parametron within a single period of its oscillation. A parametron is a binary logic element based on a driven nonlinear resonator. It features two stable phase states that define an artificial spin. The most basic operation performed on a parametron is a bit flip between these two states. 
 Thus far, this operation involved changing the energetic population of the resonator and therefore required a number of oscillations on the order of the quality factor $Q$. Our technique takes a radically different approach and relies on rapid control of the underlying potential. 
 Our work represents a paradigm shift for phase-encoded logic operations by boosting the speed of a parametron bit flip to its ultimate limit.


\end{abstract}


\maketitle


\paragraph{Introduction.}
Since the invention of the solid-state transistor, the overwhelming majority of computers followed the von Neumann architecture that strictly separates logic operations and memory~\cite{Neumann_1993,Godfrey_1993}. 
Today, there is a revived interest in alternative computation models accompanied by the necessity to develop corresponding hardware architectures~\cite{Kirkpatrick_1983, Hoppensteadt_1999, Georgescu_2014, Carleo_2017}.
For example, phase-based logic architectures can be realized with artificial spins such as the ‘parametron’ that arises in driven nonlinear resonators~\cite{Goto_1959, Neumann_1959, Sterzer_1959, Hosoya_1991, Mahboob_2011, Wang_2013, Mahboob_2016, Inagaki_2016, Goto_2016, Csaba_2016, Puri_2017}. The parametron encodes binary information in the phase state of its oscillation. 
It enables, in principle, logic operations without energy transfer and the corresponding speed limitations~\cite{Roychowdhury_2015}.


        \begin{figure}[b]
        \includegraphics[width=\columnwidth]{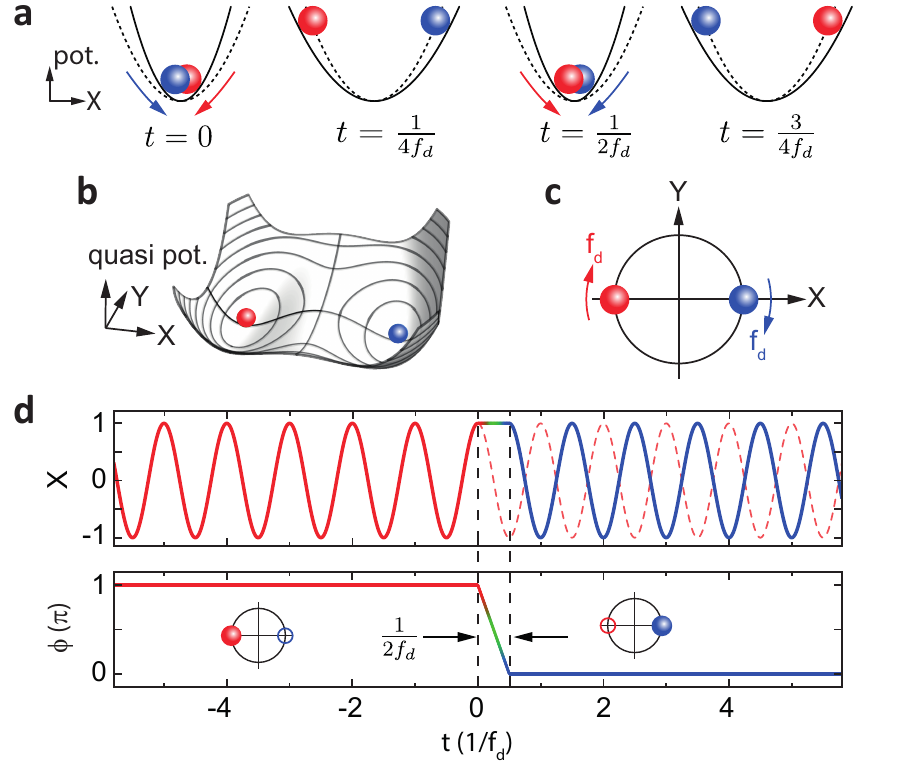}
        \caption{\label{fig:fig1} Parametron phase states and basic idea of rapid phase flipping.
        (a)~Parametric driving corresponds to a harmonic modulation of the resonator's natural frequency $f_0$.    
        Solid (dashed) lines represent the modulated (original) potential. If the drive is sufficiently strong, the resonator locks to $f_d$ and settles into one of two stable phase states that are separated by $\pi$, illustrated as oscillating red and blue spheres.
        (b)~In phase space, the parametrically driven resonator experiences an effective double-well potential, which is the key signature of the parametron. The phase states now appear as stationary red and blue spheres in the quasi-potential minima.
        (c)~Simplified illustration of the parametron in phase space. In the lab frame, the two states rotate around the origin at frequency $f_d$.
        (d)~Illustration of rapid phase flip. The parametron is initialized in the red phase state ($\phi=\pi$). At time $t=0$, the phase evolution of the system is paused for half an oscillation period by freezing the resonator's position. Upon release, the parametron resumes oscillation in the blue phase state ($\phi=0$).}
        \end{figure}

The parametron is a logic device employing the principle of parametric driving~\cite{Dykman_1998, Lifshitz_Cross, Mahboob_2008, Wilson_2010, DykmanBook, Mahboob_2011, Lin_2014}. Consider a resonator whose natural frequency $f_0$  is modulated at a drive frequency $2f_d$. If $f_d$ is chosen close to $f_0$, and the modulation is sufficiently strong, the resonator experiences a negative effective damping and is forced to oscillate at $f_d$ with large amplitude, as illustrated in Fig.~\ref{fig:fig1}(a). With the frequency of the motion being half that of the modulation, the resonator undergoes a spontaneous time-translation symmetry breaking~\cite{Wilczek2012, Leuch_2016}. As a result, the system is locked to one of the two available phase states that are degenerate in amplitude but separated by $\pi$ in phase (relative to a clock running at $f_d$). 
In phase space spanned by normalized displacement $X$ and momentum $Y$~\cite{Rabi_1954}, this locking mechanism can be illustrated by the quasi-potential landscape shown in Fig.~\ref{fig:fig1}(b). The quasi-potential features a double-well structure, where each well corresponds to a stable phase state. The two phase states of the parametron represent a classical bit or, analogously, an Ising spin. In the lab frame, the states rotate around the phase-space origin at the drive frequency $f_d$ [Fig.~\ref{fig:fig1}(c)].


While the parametron was already patented at the dawn of the digital era~\cite{Goto_1959, Neumann_1959}, it is only with recent experimental advances that an implementation of the concept appears useful. Research groups using nanomechanical resonators, Josephson junction circuits, and optical parametric oscillators have devised prototypical parametron-based Ising machines that may solve NP-hard problems much faster than conventional computers~\cite{Lin_2014, Marandi_2014, Mahboob_2016, Goto_2016, Inagaki_2016, Puri_2017}. 
The most basic logic operation on a parametron is a bit flip, corresponding to a phase change of $\pi$ of the underlying resonator. Thus far, parametrons have been flipped by first depleting the resonator and then re-energizing it in the opposite phase state~\cite{Mahboob_2008,Mahboob_2016}.
The flipping speed of this method is limited by the ring-down time $\tau= Q/(\pi f_0)$, where $Q\gg 1$ is the quality factor of the resonator. This speed limitation is directly related to the energy gap between energized and depleted states. However, flipping the phase state of a parametron does not strictly require energy transfer. Indeed, the two logic states are degenerate in energy and protected by a `phase gap'~\cite{Roychowdhury_2015}.
It should therefore be possible to devise a protocol to flip between the phase states at a speed much faster than the ringdown time $\tau$, which is often (erroneously) deemed a fundamental limit for resonator operations~\cite{Budakian_2006, Liu_2008}.
Despite the fact that such a protocol would unlock the full potential of phase-encoded logic, an experimental demonstration has remained elusive to date.



In this paper, we experimentally demonstrate flipping between the two phase states of a parametron within a single oscillation period. Our technique allows logic operations on a time-scale of $1/f_0$, and therefore $Q$ times faster than the ring-down time. Our protocol temporarily freezes (or slows down) the evolution of a resonator to bridge the phase gap separating its phase states. 
The speed of our method relies on the fact that it does not require energy transfer into or out of the system. The demonstrated protocols are platform independent.
We present two complementary variations of our phase-flip paradigm on different experimental systems and assess their performances.
Our results call for a reevaluation of the fundamental limits for high-speed and low-energy computation using parametron bits.

        \begin{figure*}
        \includegraphics[width=\textwidth]{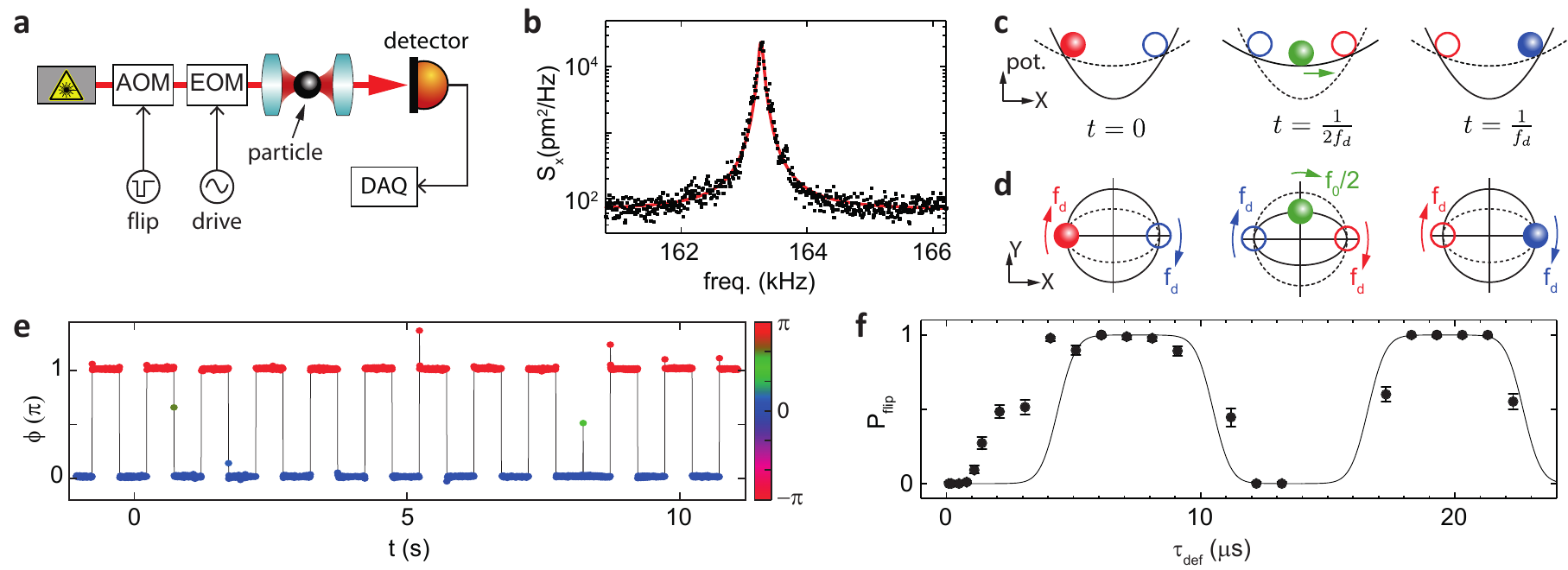}
        \caption{\label{fig:fig2} Experimental demonstration of phase flip via potential deformation.
        (a)~Experimental setup. A silica nanoparticle (diameter 136~nm) is trapped in a focused laser beam (wavelength 1064~nm) inside a vacuum chamber (not shown). The stiffness of the optical potential can be modulated with an electro-optic modulator (EOM). The particle displacement is detected with a quadrant photo diode (QPD).
        (b)~Thermally driven power spectral density $S_x$ of the particle displacement. From the red line fit, we extract a quality factor $Q = 1970$.
        (c)~Schematic illustration of the phase-flip protocol. The parametron is initialized in the red phase state. When the particle reaches its maximum displacement, we reduce the resonator frequency from $f_0$ (potential sketched as solid line) to $f_0/2$ (dashed line) by attenuating the laser intensity with an acousto-optic modulator (AOM). We then let the particle evolve for the pulse length $\tau_\text{def}=1/f_0$, such that the phase states of the parametron undergo a full oscillation, while the particle only traverses the trap and acquires a phase delay of $\pi$.
        (d)~Same as (c) but illustrated in phase space.
        (e)~Measured phase of the parametron as a function of time. A switch of the potential as outlined in (c) and (d) is applied at a rate of 2~Hz with $\tau_\mathrm{def}=8.1~\mu$s, periodically flipping the parametron phase state. Note the failed flip around 8~s.
        (f)~Flipping probability $P_\text{flip}$ for varying pulse length $\tau_\text{def}$. Our model (black line) takes into account the finite thermal population of the resonator (see Supplemental Material). Error bars represent statistical uncertainty.}
        \end{figure*}



\paragraph{Phase-flip protocols.}
The general idea for rapid parametron phase flipping is illustrated in Fig.~\ref{fig:fig1}(d). 
The resonator is initially in one of the two stable phase states. Without limitation of generality, let us consider the red phase state with phase $\phi=\pi$.
At $t=0$, the resonator evolution is frozen (or slowed down), such that it acquires a phase delay relative to its initial state. Careful timing results in a delay of exactly $\pi$. Upon release, the resonator resumes oscillation in the blue phase state with phase $0$. In the following, we consider two methods to achieve such a phase delay by $\pi$. They make use of `potential deformation' and `potential displacement', corresponding to a change in the restoring force and to the application of an external force, respectively. 

\paragraph{Phase flip via potential deformation.}
We first demonstrate rapid parametron phase flipping via potential deformation, corresponding to switching the underlying resonator's natural frequency $f_0$. As an experimental platform, we use a silica nanoparticle optically levitated in a focused laser beam in vacuum, as illustrated in Fig.~\ref{fig:fig2}(a) (see \cite{Gieseler_2012} and Supplemental Material for details). The light scattered by the particle provides us with a measurement of its position. Each degree of freedom of the particle's center-of-mass represents a nonlinear resonator~\cite{JanNature}. To minimize the effect of thermal fluctuations, we feedback-cool all three degrees of freedom to a temperature of 1~K. Throughout this work, we focus on a single oscillation mode with a resonance frequency $f_0 \sim 164$~kHz. The power spectral density of the feedback-cooled mode under consideration is shown in Fig.~\ref{fig:fig2}(b).

Weak periodic modulation of the trapping laser intensity turns the levitated particle into a parametron. In contrast, a sudden and strong reduction of the laser intensity leads to a deformation of the potential and can be used for phase flips. Consider the particle confined in a potential of natural frequency $f_0$ under parametric driving at $2f_d$ [with $f_d\sim f_0$), such that the parametron is locked to one of the two stable phase states (Figs.~\ref{fig:fig2}(c-d)]. When the particle reaches its maximum displacement (and its velocity vanishes), we reduce the power of the trapping laser to switch the natural oscillation frequency to $f_0/2$ for a time $\tau_\text{def}$. If we choose $\tau_\text{def} = 1/f_0$, the particle has time to travel to the opposite side of the potential. At this moment, we switch the laser intensity (and thus the trap stiffness) back to its original value and the particle continues to oscillate at a frequency $f_d$. Importantly, relative to the clock at $f_d$, the phase state of the parametron has been flipped by $\pi$ during the protocol.

We show an experimental demonstration of our idea in Fig.~\ref{fig:fig2}(e), where we plot the phase state of the optically levitated parametron as a function of time. The measurement signal is the output of a lock-in amplifier fed with the position signal. The trap frequency is switched twice per second from $f_0 = 164$~kHz to 82~kHz for a duration $\tau_\text{def} = 8.1~\mu$s.  Indeed, we observe two phase states separated by $\pi$ and flipping between them at the expected rate of 2~Hz. The phase flips happen instantaneously on the timescale set by the 220~Hz bandwidth of our lock-in detection.

A striking feature in Fig.~\ref{fig:fig2}(e) is the failed phase flip around 8~s, indicating that the success probability $P_\text{flip}$ of our potential deformation scheme is less than unity (we define $P_\text{flip}$ as the ratio of observed phase flips to flipping attempts).
We attribute this observation to the fact that we did not choose the nominally ideal value of $\tau_\text{def} = 6.1~\mu$s.
To corroborate this hypothesis, we record $P_\text{flip}$ for varying $\tau_\text{def}$. In Fig.~\ref{fig:fig2}(f), we observe that $P_\text{flip}$ is indeed a periodic function of $\tau_\text{def}$ with the expected period $2/f_d$. When $\tau_\text{def}$ is an even multiple of $1/f_d$, the parametron phase remains unaltered by the pulse and $P_\text{flip}$ vanishes. In contrast, for $\tau_\text{def}$ equal to an odd multiple of $1/f_d$, $P_\text{flip}$ approaches unity. We note that a pulse of length $\tau_\text{def}=2/f_d$ can be interpreted as a sequence of two back-to-back pulses of length $1/f_d$. The data in Fig.~\ref{fig:fig2}(f) therefore demonstrate that it is possible to fully exploit the switching speed of our method by concatenating several rapid phase-flips.

Figure~2(f) reveals that the transitions of $P_\text{flip}(\tau_\text{def})$ between zero and unity are not infinitely sharp but display a finite width of about 2~$\mu$s, which we attribute to thermomechanical fluctuations.
The solid line in Fig.~\ref{fig:fig2}(f) indicates a model calculation of $P_\text{flip}$ based on thermal phase noise (see Supplemental Material). This model reproduces our data well for $\tau_\text{def} > 5~\mu$s. We attribute the deviations between data and model for short $\tau_\text{def}$ to the finite response time and the resulting transients of the modulator that switches the laser power.

We note that in our experiment, we triggered a phase flip when the resonator displacement was at its maximum. The protocol is, however, applicable with any starting condition (see Supplemental Material). Indeed, under the applied potential deformation, a harmonic oscillator with initial phase state $(X,Y)$
will always evolve towards $-(X,Y)$
within half a period. By extension, the protocol is applicable to arbitrary mixtures of states, including thermal states.
Finally, we point out that the flipping time of our protocol could be further reduced to $1/(2f_d)$ by completely turning off the trapping potential. 
For our particular experimental situation, switching off the potential also reduces the parametric drive to zero. On the short time scale of the flipping process, this is not problematic because parametric locking is only effective on time scales of the order of $\tau$.
However, the scheme implemented in this work is significantly more resilient against inevitable thermal fluctuations of the particle motion which can lead to particle loss.



        \begin{figure*}
        \includegraphics[width=\textwidth]{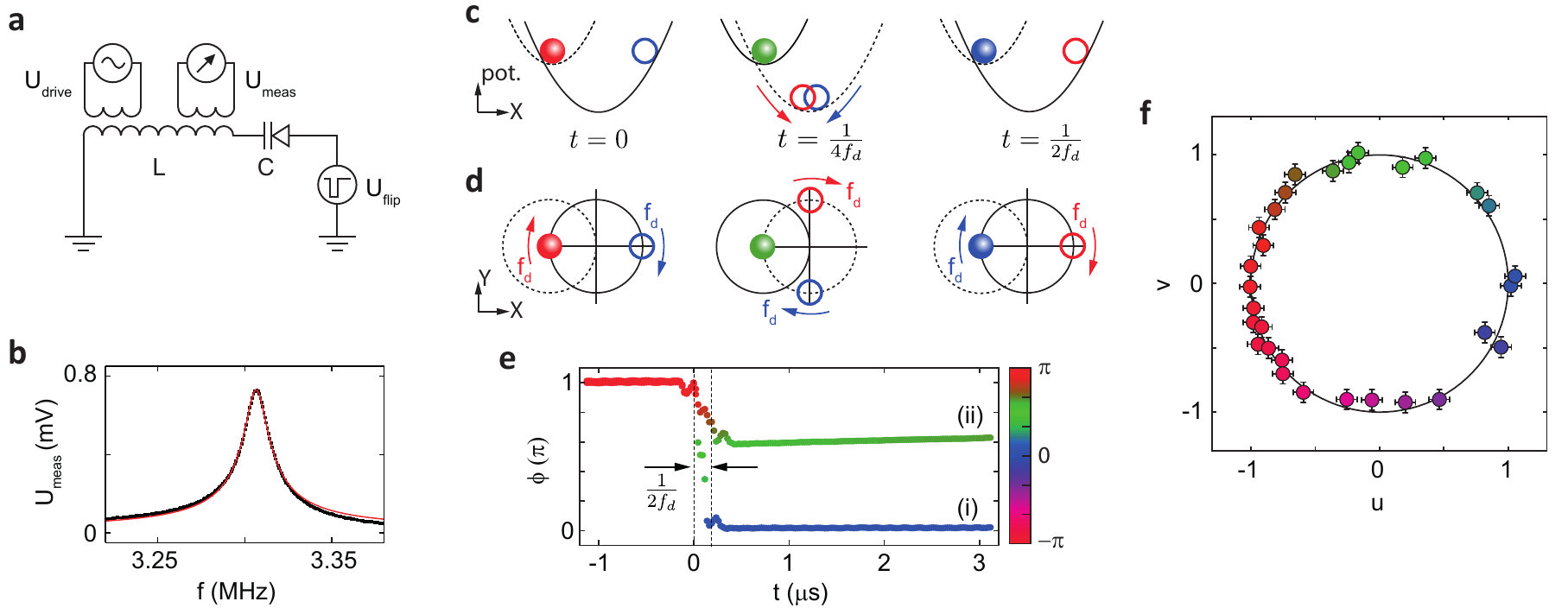}
        \caption{\label{fig:fig3} Experimental demonstration of phase flips via potential displacement.
        (a)~Schematic of the electrical LC resonator circuit with a varicap diode to provide a nonlinear capacitance $C$. 
        (b)~Linear response of the resonator to a small external driving voltage ($U_{\rm{drive}} = 50~$mV). From the red line fit we extract $f_0 = 3.3~$MHz and $Q = 245$.
        (c)~Illustration of the phase-flip protocol. The parametron is initialized in the red phase state. When the resonator reaches its maximum displacement at $t=0$, the potential is displaced by an external force (from dashed to solid lines) such that the resonator is momentarily at rest. At $t = 1/(2f_d)$, the force is turned off and the parametron resumes its evolution, now in the blue phase state.
        (d)~Same as (c) but illustrated in phase space.
        (e)~Demonstration of two different phase flips, performed with (i) $\tau_\text{dis} = 153~$ns, the ideal pulse duration for flipping, and with (ii) $\tau_\text{dis} = 230~$ns. The signal was demodulated by a digital lock-in amplifier and filtered for clarity (see Supplemental Material).
        (f)~Results of flipping experiments with varying $\tau_\text{dis}$. 
        Each datapoint represents the state of the resonator directly after a pulse. Here, $u=X\cos(2\pi f_d t) - Y\sin(2\pi f_d t)$ and $v=Y\cos(2\pi f_d t) + X\sin(2\pi f_d t)$ are the phase-space quadratures in a frame rotating at the drive frequency $f_d$. A black circle serves as a guide to the eye.
        }
        \end{figure*}

\paragraph{Phase flip via potential displacement.}
In the following, we demonstrate that rapid parametron phase flips are also possible by displacing the potential, corresponding to the application of a force to the resonator. We experimentally realize this method with the electrical LC circuit illustrated in Fig.~\ref{fig:fig3}(a) (see \cite{Nosan_2019} and Supplemental Material for details). Here, the resonator displacement corresponds to the charge separated across the varicap diode with capacitance $C$, and the role of the force is assumed by a voltage $U_\text{flip}$. We characterize our resonator in the linear regime by applying a weak drive tone $U_\text{drive}$ whose frequency we sweep around $f_0$ while recording the output voltage $U_\text{meas}$, as shown in Fig.~\ref{fig:fig3}(b). The circuit becomes a parametron under sufficiently strong driving close to $2f_0$.

We use this system to realize the phase-flipping scheme detailed in Figs.~\ref{fig:fig3}(c-d). When the resonator displacement reaches its maximum value, a force is applied to counter the restoring force and to freeze the resonator evolution. This equals a displacement of the potential by the oscillation amplitude, such that the resonator temporarily finds itself at the potential center. After the force is turned off, the resonator has acquired a phase delay of $\pi$ relative to its original evolution and is stable in the opposite phase state.

In Fig.~\ref{fig:fig3}(e), we show two examples of the behavior of the system for different pulse lengths $\tau_\text{dis}$. In the first example, the pulse length is set to $\tau_\text{dis}=1/(2f_0)$, the ideal pulse length for a bit flip. Indeed, the parametron flips its phase state by $\pi$ (i, blue data points). In the second example, we set $\tau_\text{dis}=1.5\times 1/(2f_0)$ (ii, green). Here, the parametron is transferred into a state between the two stable phase states and evolves towards one of them on a timescale given by $Q/f_0 \sim 74~\mu$s after the flip.


In Fig.~3f, we plot the state of the resonator at $t = 0.7~\mu$s after the start of a bit flip in phase space (in a frame rotating at the drive frequency) for different values of $\tau_\text{dis}$. The amplitude of the parametron after the flipping protocol (corresponding to the radial distance from the plot center) is independent of $\tau_\text{dis}$, which results from the fact that the resonator's evolution is frozen at the point of maximum displacement and vanishing velocity. Our data demonstrates that via the choice of $\tau_\text{dis}$ we can transfer the parametron to any point on the unit circle in phase space, in particular to the two stable phase states.


\paragraph{Discussion and Conclusion.} The two experimental demonstrations in Figs.~\ref{fig:fig2} and \ref{fig:fig3} establish a new paradigm for resonator-based logic operations. Parametron phase flips can be achieved within a single oscillation period and completely independently from the quality factor $Q$. This finding opens up new possibilities for applications that use parametrons as phase logic units~\cite{Goto_1959, Neumann_1959, Sterzer_1959, Mahboob_2011, Wang_2013, Mahboob_2016, Inagaki_2016, Csaba_2016, Puri_2017, Puri_2017_NC, Yamamura_2017, Zhao_2018, Goto_2018}. The states of the parametrons may be initialized and flipped irrespective of the (desirable) high quality factors of the underlying resonators, and the flips do not necessarily involve energy exchange with a bath. In this way, our schemes reconcile the two seemingly disparate notions of rapid logic operations and long state coherence~\cite{Tsaturyan_2017, Ghadimi_2018}. Beyond computation, rapid phase flips allow encoding binary information through phase-shift keying~\cite{RappaportBook}. 
While current phase-shift keying techniques use an oscillator with constant amplitude and phase and achieve different phase-space states through post-processing, our demonstrations show that information encoding on the level of the resonator itself is feasible. 
This may enable ultra-compact and low-power encoders for specialized applications such as autonomous nanobots in medical research~\cite{Duncan_2011, Li_2017}.

There are several factors that significantly relax the required conditions for large-scale implementations of our technique. First, the symmetry protection of the parametron makes the phase-flips very stable in the presence of phase noise~\cite{Roychowdhury_2015}. Consecutive rapid flips result in the correct state as long as the summed phase error is below $\pi/2$. After a sequence of rapid flips, phase errors will self-correct through relaxation within the double-well. Second, the external parametric driving signal can be utilized as a clock with large signal-to-noise ratio. Estimating the momentary state of a parametron is thus fault tolerant up to $\pi/2$, while the amplitude is generally known.

The physics explored within our work may be translated to nonlinear high-frequency resonators based on Josephson junction circuits~\cite{Wilson_2010, Lin_2014, Puri_2017, DykmanBook}, micro- and nanomechanical resonators such as carbon nanotube and graphene devices~\cite{Rugar_1991, Mahboob_2008, Karabalin_2010, Villanueva_2011, Eichler_2011_NL, Mathew_2016, DykmanBook}, optical parametric oscillators in nonlinear media~\cite{Wang_2013, Marandi_2014, Inagaki_2016}, trapped ions~\cite{Ding_2017}, and cold atom lattices~\cite{Schilke_2012}. It is thus a highly general concept that is potentially useful in a wide variety of experiments and future applications.

See Supplemental Material [URL to be inserted] for experimental details, a model of the bit flip success rate, as well as Fokker-Planck numerical simulations, which includes Refs.~\cite{Gardiner_2009, Gieseler_2014, Jain_2016}.


\begin{acknowledgments}
We are indebted to Peter M\"arki, Nils Hauff, David Ruffieux and Can Knaut for valuable discussions and technical assistance during this project. This research was supported by ERC-QMES (Grant No. 338763), the NCCR-QSIT program (Grant No. 51NF40-160591), the Swiss National Science Foundation (CRSII5 177198/1, PP00P2\textunderscore 163818), the Michael Kohn Foundation, the ETH Z\"urich Foundation, and a Public Scholarship of the Development, Disability and Maintenance Fund of the Republic of Slovenia (11010-247/2017-12). 
\end{acknowledgments}


%
\end{thebibliography}%


\end{document}